# Solvable Schrodinger Equations of Shape Invariant Potentials Having Superpotential $W(x,A,B) = A\tanh(px) + B\tanh(6px)$

Jamal Benbourenane [1], Mohamed Benbourenane [1], Hichem Eleuch [2,3]

[1] Department of Mathematics and Physical Sciences,
California University, PA, USA

[2] Department of Applied Physics and Astronomy, University
of Sharjah, UAE

[3] Institute for Quantum Science and Engineering, Texas A&M,
College Station, Texas, USA

A new proposed one dimensional time independent Schrödinger equation is solved completely using shape invariance method. The corresponding potential is given by
$$V_-(x,A,B,p) = -A(\operatorname{sech} px)^2 - 6Bp(\operatorname{sech} 6px)^2 + (\tanh px - 6\tanh 6px)^2$$
with superpotential $W(x,A,B) = A\tanh(px) + B\tanh(6px)$.

We derive the exact solutions of the family of Schrödinger equations with the $V_-$ potential partner using supersymmetric quantum mechanics technique of a superpotential having shape invariance property, and where the discrete spectrum and the corresponding eigenfunctions are determined exactly and in closed form.

It is well-known that Schrödinger equations are challenging to solve in closed form, and only a few of them are known. Finding new equations with exact solutions is crucial in understanding the hidden physical properties near turning points where numerical methods fail in these vicinities. This result has potential applications in nuclear physics and chemistry where the antagonist forces have a prominent presence.

## 1  Introduction

Schrödinger equation was developed (1926) [50] by Erwin Schrödinger to study the atomic spectra. Schrödinger equation is becoming a fundamental equation in modern science beyond atomic physics, nuclear physics, molecular chemistry, nanostructures, quantum processing, etc. [29] [39]. It is used for studying quantum biology phenomena such as photosynthetic reactions and DNA [38] [42]. Its application reached other exotic fields like quantum finance, where scientists speak about Hamiltonian options and quantum finance [3]. The Schrödinger equation's eigenvalues for a defined potential are conjectured to be the Riemann-zeta function's zeros [9] [8], which can link prime numbers with the eigenvalues of the Hamiltonian. This leads to cryptography field which is linked to the security of our connected digital world. Therefore, finding the solutions of the Schrödinger equation's eigenvalue problem is crucial for all these fields of science. Several methods such as WKB, SWKB, perturbation theory, variational method, and ERS technique allow us to derive the Schrödinger equation's approximate solutions [12] [28] [54]. However, these solutions may not detect some of the hidden important physical properties of the quantum system. Exact solutions are the utmost goal for investigating such properties.

In the literature we can find only few potentials with exact solutions where the eigenvalues and eigenfunctions are given explicitly in closed forms and these potentials were discovered in the first half of last century starting by Schrödinger with the harmonic oscillator [51], Coulomb, 3-D oscillator, Morse [43] (1929), Eckart [21] (1930), Rosen-Morse I&II [47] (1932), Pöschl-Teller I&II [46],(1933), Scarf I&II [49] (1958). Recently, we have added two new potentials to the list [7].

Supersymmetry (SUSY) is among the most successful methods for evaluating physical systems. This technique was introduced by Nicolai and Witten in non-relativistic quantum mechanics [44] [55].

In this paper, we present the exact spectrum and eigenfunctions of the potential in the form

$$V_-(x,B) = -\frac{1}{6}(B-6p)(\mathrm{sech} px)^2 - 6Bp(\mathrm{sech} 6px)^2 + (-\frac{1}{6}(B-6p)\tanh px + B\tanh 6px)^2$$

with superpotential given in the form

$$W(x,B,p) = -\frac{1}{6}(B-6p)\tanh px + B\tanh 6px.$$

In the next section, we introduce the concept of SUSY. Section 3 is devoted to shape invariance property that allows generating exact solutions. In section 4, we will completely solve the new potential using the SUSY method. The potential satisfies the shape invariance property. Therefore, we evaluate its eigenvalues and eigenfunctions in closed form. In section 5, we compare the known numerical methods, namely, shooting method, WKB method, SWKB method, and NDEigensystem method in Mathematica, to our exact solutions. We conclude by summarizing our discovery and its implications in section 6.

## 2 Supersymmetry

Given a potential $V_-(x)$, we seek to built a potential partner $V_+(x)$, where these two potentials have the same energy eigenvalues, except for the ground state.

These potential partners are defined by

$$V_\pm(x) = W^2(x) \pm W'(x) \tag{1}$$

where $W(x)$ is called the superpotential.

The associated Hamiltonian to these potential partners are given by

$$H_- = A^\dagger A, \quad H_+ = AA^\dagger \tag{2}$$

where

$$A^\dagger = -\frac{d}{dx} + W(x), \quad A = \frac{d}{dx} + W(x) \tag{3}$$

The one dimensional time-independent Schrödinger equation with eigenstate $E$ and potential $V(x)$ is defined by

$$(-\frac{d^2}{dx^2} + V(x))\Psi = E\Psi \tag{4}$$

and the two Hamiltonians associated with the Schrödinger equation are written in the form

$$H_- = -\frac{d^2}{dx^2} + V_-(x), \quad H_+ = -\frac{d^2}{dx^2} + V_+(x), \tag{5}$$

These two Hamiltonians (5) have their eigenvalues (energy levels) and their eigenfunctions (wavefunctions) intertwined. That is if $E_0^{(-)}$ is an eigenstate of the Hamiltonian $H$ and its associated eigenfunction $\Psi_0^{(-)}$, then the Hamiltonian $H_+$ will have the same eigenstate $E_0^{(-)}$ and its eigenfunction is given by $A\Psi_0^{(-)}$, and vice-versa if we change $H_+$ by $H_-$.

The two Hamiltonians are both positive semi-definite operators, so their energies are greater than or equal to zero.

For the Hamiltonian $H_-$, we have

$$H_-\Psi_0^{(-)} = A^\dagger A\Psi_0^{(-)} = E_0^{(-)}\Psi_0^{(-)} \tag{6}$$

So, by multiplying on the left side of equation (6) by the operator $A$, we obtain

$$AA^\dagger A\Psi_0^{(-)} = E_0^{(-)}(A\Psi_0^{(-)}),$$

so that

$$H_+(A\Psi_0^{(-)}) = E_0^{(-)}(A\Psi_0^{(-)}).$$

Similarly, for $H_+$

$$H_+\Psi_0^{(+)} = AA^\dagger\Psi_0^{(+)} = E_0^{(+)}\Psi_0^{(+)} \tag{7}$$

and multiplying the left side of (7) by $A^\dagger$ we obtain

$$H_-(A^\dagger\Psi_0^{(+)}) = E_0^{(+)}(A^\dagger\Psi_0^{(+)}). \tag{8}$$

The two Hamiltonians' eigenfunctions and their exact relationships will depend on whether the quantity $A\Psi_0^{(-)}$ is zero or nonzero. If $E_0^{(-)}$ is zero (nonzero), it means an unbroken (broken) supersymmetric system. For a full discussion of this cases, see the references [52] [53].

Thus, here we will consider only the case of unbroken supersymmetry, where $A\Psi_0^{(-)} = 0$. In this case, this state has no SUSY partner since the ground state wavefunction of $H_-$ is annihilated by the operator $A$, and in which case $E_0^{(-)} = 0$.

It is then clear that the eigenstates and eigenfunctions of the two Hamiltonians $H_-$ and $H_+$ are related by (for $n = 0,1,2,...$)

$$E_0^{(-)} = 0, \quad E_n^{(+)} = E_{n+1}^{(-)}, \tag{9}$$

$$\Psi_n^{(+)} = \left(E_{n+1}^{(-)}\right)^{-1/2} A\Psi_{n+1}^{(-)} \tag{10}$$

$$\Psi_{n+1}^{(-)} = \left(E_n^{(+)}\right)^{-1/2} A^\dagger\Psi_n^{(+)}. \tag{11}$$

This process has some similarities to the harmonic oscillator by applying the creation and annihilation operators.

By knowing $W(x)$, then the ground state wavefunction $\Psi_0^{(-)}$ can be expressed by

$$\Psi_0^{(-)} = N \; e^{-\int W(x)dx} \tag{12}$$

where $N$ is the normalized constant, while by knowing the ground state wavefunction $\Psi_0^{(-)}$ the superpotential $W$ can be expressed in the form

$$W(x) = -\frac{d}{dx}\log\left(\Psi_0^{(-)}(x)\right) \tag{13}$$

Note that $\Psi_0^{(-)}$ is a solution to the Schrödinger Eq. (4) with potential $V$, that is,

$$\Psi_0 = \Psi_0^{(-)} = N \; e^{-\int W(x)dx}$$

We can see that by normalizing the eigenfunction $\Psi_n$ of $H$, the wavefunction $\Psi_{n+1}$ is also normalized. Also, the operator $A$ (as well as $A^\dagger$) converts an eigenfunction of $H_-$ ($H_+$) into an eigenfunction of $H_+$ ($H_-$) with the same energy.

The bound state wavefunctions must converge to zero at the two ends of its interval domain, therefore the two statements $A\Psi_0^{(-)} = 0$ and $A^\dagger\Psi_0^{(+)} = 0$ cannot be satisfied at the same time, and so only one of the two ground state energies, $E_0^{(-)}$ and $E_0^{(+)}$, can be zero, while the other bound state energy will be positive. So, we will consider by convention $W$ such that $\Psi_0$ is normalized with

$$E_0^{(-)} = 0, \; E_1^{(-)} = E_0^{(+)} > 0. \tag{14}$$

## 3  Shape Invariance

As we have seen in the last section, in order to solve the Schrödinger equation (4) for a potential $V(x)$ using the method of supersymmetry, we need to construct the superpotential $W$. This is equivalent to solving the Riccati differential equation (1). However, Riccati equation is also an unsolved equation except for a limited number of cases. Nevertheless, Riccati equation happens to be more tractable then the original Schrödinger equation for some potentials with a particular geometric property of the shape of the potential partners as it can be seen in the known solvable potentials [51] [43] [21] [47] [46] [49] [7] [6].

We will define potentials to be shape invariant if their dependence on $x$ is similar and they only differ on some parameters appearing in their expressions. This similarity is described by the following relation :

$$V_-(x, a_1) + h(a_1) = V_+(x, a_0) + h(a_0) \tag{15}$$

where the parameter $a_1$ depends on $a_0$, i.e. $a_1 = f(a_0)$, and then, $a_2 = f(a_1) = f^2(a_0)$, and by recurrence $a_k = f^k(a_0)$ where $a_0 \in \mathbb{R}^m$, and $f: \mathbb{R}^m \to \mathbb{R}^m$, is called a parameter change function, see [1] [11] [14] [17] [26].

With the shape invariance property, the superpotentials parameters change. However, the potential partners have similar shapes and only differ by a non-zero constant.

The condition of shape invariance (15) also proved to be another major hurdle on finding its solutions. Since only the already known solvable potentials, namely, harmonic oscillator,

Coulomb, 3D-oscillator, Morse, Rosen-Morse I &II, Eckart, Scarf I&II, Pőschl-Teller I&II, have been shown to satisfy this condition. In our last paper, we introduced two newly discovered potentials that satisfy the shape invariance property [7] and have derived explicit exact eigenvalues and eigenfunctions. We would like to mention here that a shape invariant potential was proposed in a series form, see [15], however, this potential is not in a closed form.

More precisely, we aim to determine all the bound state energies and the expression of their wavefunctions.

From now on, we will consider the potential $V(x, a)$ defined in the Schrödinger equation (4) as a shape-invariant potential. Therefore, the two potential partners $V_-(x, a_1)$ and $V_+(x, a_0)$ have the same dependence on $x$, up to the change in their parameters, and their Hamiltonians $H_-(x, a_1)$ and $H_+(x, a_0)$ differ only by a vertical shift given by $C(a_0) = h(a_1) - h(a_0)$,

$$V_+(x, a_0) = V_-(x, a_1) + C(a_0) \tag{16}$$

where the potential partners are defined by

$$V_-(x, a_1) = W^2(x, a_1) - W'(x, a_1), \tag{17}$$
$$V_+(x, a_0) = W^2(x, a_0) + W'(x, a_0) \tag{18}$$

The zero-ground state wavefunction is given by

$$\Psi_0^{(-)}(x, a_0) \propto e^{-\int W(x, a_0) dx}. \tag{19}$$

By differentiating twice, we can show that this wavefunction satisfies the Schrödinger equation

$$-\Psi_0''(x) + V(x)\Psi_0(x) = 0 \tag{20}$$

with the potential $V(x) = V_-(x, a_0)$ and the associated eigenvalue $E_0^{(-)} = 0$.

The first excited state $\Psi_1^{(-)}$ of $H_-(x, a_1)$ is given here, where we omit the normalization constant,

$$\Psi_1^{(-)}(x, a_0) = A^\dagger(x, a_0)\Psi_0^{(+)}(x, a_0) = A^\dagger(x, a_0)\Psi_0^{(-)}(x, a_1). \tag{21}$$

The eigenvalue associated with this Hamiltonian has the following expression

$$E_1^{(-)} = C(a_0) = h(a_1) - h(a_0) \tag{22}$$

The eigenvalues of the two Hamiltonians $H_+$ and $H_-$ have the same eigenvalues except for the additional zero-energy eigenvalue of the lower ladder Hamiltonian $H_-$. They are related by

$$E_0^{(-)} = 0, \quad E_{n+1}^{(-)} = E_n^{(+)}, \tag{23}$$
$$\Psi_n^{(+)} \propto A\Psi_{n+1}^{(-)}, A^\dagger \Psi_n^{(+)} \propto \Psi_{n+1}^{(-)}, \quad n = 0,1,2,\ldots \tag{24}$$

where we have iterated this procedure to construct a hierarchy of Hamiltonians

$$H_\pm^{(n)} = -\frac{d^2}{dx^2} + V_\pm(x, a_n) + \sum_{k=0}^{n-1} C(a_k) \tag{25}$$

and then derive the $n^{th}$ excited eigenfunction and eigenvalues by

$$\Psi_n^{(-)}(x, a_0) \propto A^\dagger(x, a_0) A^\dagger(x, a_1) \ldots A^\dagger(x, a_n) \Psi_0^{(-)}(x, a_n) \tag{26}$$

$$\begin{aligned} E_0^{(-)} &= 0, \\ E_n^{(-)} &= \sum_{k=0}^{n-1} C(a_k) = \sum_{k=0}^{n-1} h(a_{k+1}) - h(a_k) \\ &= h(a_n) - h(a_0), \text{ for } n \geq 1. \end{aligned} \tag{27}$$

where $a_k = f(f(\ldots f(a_0))) = f^k(a_0)$, $k = 0, 1, 2, \ldots, n-1$.

Therefore, by knowing the superpotential not only we know the potential but also its ground state. From the algorithm above, the whole spectrum of the Hamiltonian $H_-$ ($H_+$) can be derived by the supersymmetry quantum mechanics method.

## 4 Exactly solvable potentials with superpotential $W(x, A, B) = A\tanh px + B\tanh 6px$

We consider the superpotential
$$W(x, A, B) = A\tanh px + B\tanh 6px \tag{28}$$
with its potential partners satisfying the shape invariance condition (15), we obtain

$$V_-(x, A_1, B_1) = -pA_1 \text{sech}^2 px - 6pB_1 \text{sech}^2 6px + (A_1 \tanh px + B_1 \tanh 6px)$$
$$V_+(x, A_0, B_0) = pA_0 \text{sech}^2 px + 6pB_0 \text{sech}^2 6px + (A_0 \tanh px + B_0 \tanh 6px)$$

which can be rewritten in a more simplified form, by using the sequence defined in (32) (33) (34), and which clearly exhibit the shape invariance property,

$$V_-(x, B_1) = -\frac{1}{6}(35B_1 - 6p)p + \frac{1}{36} B_1(B_1 + 6p) U(x), \tag{29}$$

$$V_+(x, B_0) = \frac{1}{6}(35B_0 + 6p)p + \frac{1}{36} B_0(B_0 - 6p) U(x), \tag{30}$$

as the two potentials look similar except in the parameters appearing in them, and where $U(x)$ is the nonnegative function defined by
$$U(x) = (\tanh px - 6\tanh 6px)^2. \tag{31}$$

We have defined the sequence $a_k = (A_k, B_k)$ of the shape invariance property as follows:

$$A_0 = -\frac{B_0}{6} + p, \tag{32}$$

$$A_k = A_0 - kp, \tag{33}$$

$$B_k = B_0 - 6kp \text{ for } k = 0, 1, 2, \ldots \tag{34}$$

Since $A_k$ depend on $B_0$, all our hierarchical potentials depend therefore on $B_k$,

We can see that the value
$$(V_-)_{\min} = -\frac{1}{6}(35B_0 + 6p)p \tag{35}$$
is the minimum of $V_-(x)$ which is attained at $x = 0$.

The main potential in (20) of the Schrödinger equation $V_-(x) = V_-(x, B_0)$ is given by

$$V_-(x) = -A_0 (\text{sech} px)^2 - 6B_0 p (\text{sech} 6px)^2 + (A_0 \tanh px + B_0 \tanh 6px)^2$$

where we condensed the expression by using $A_0$ instead of $-\frac{B_0}{6}+p$ given in (32), for the sake of simplicity.

The first energy level of the potential partner $V_-$ is given by (37) by,
$$E_1^{(-)} = V_+(x,B_0) - V_-(x,B_1) = \frac{35}{6}(B_0 - 3p)p \tag{36}$$

So, we will assume $0 < p < \frac{B_0}{3}$, in order to satisfy the positive semi-definite condition (14)

$$E_1^{(-)} = E_0^{(+)} = C(B_0) = c_0^2 - c_1^2, \tag{37}$$

By recurrence, we have
$$C_k = C(B_k) = c_k^2 - c_{k+1}^2, \quad k = 0,1,2,\ldots \tag{38}$$

where,
$$c_k = (A_k + B_k)^2$$

Thus, for all $n = 0,1,2,\ldots$, the $n^{th}$ bound energy level of the potential partner $V_-$ is given by
$$E_n^{(-)} = \sum_{k=0}^{n-1} C_k = (A_0 + B_0)^2 - (A_n + B_n)^2 \tag{39}$$
$$= (A_0 + B_0)^2 - (A_0 + B_0 - 7np)^2 \tag{40}$$
$$= \left(\frac{5B_0}{6} + p\right)^2 - \left(\frac{5B_0}{2} + p - 7np\right)^2 \tag{41}$$

For $n \geq 0$, this energy can be written as
$$E_n^{(-)} = \frac{7}{3}np(5B_0 + 3(2 - 7n)p). \tag{42}$$

We also observe that the ground wavefunction expression, obtained using (19) is given by
$$\Psi_0(x,B_0) = \cosh px^{\frac{1}{2}(-1+\frac{B_0}{6p})}\cosh 6px^{-\frac{B_0}{6p}}. \tag{43}$$

The first excited state wavefunction is
$$\Psi_1(x,B_0) = \left(-\frac{d}{dx} + W(x,B_0)\right)\Psi_0(x,B_1) \tag{44}$$
$$= \frac{1}{6}(B_0 - 3p)\cosh px^{-1+\frac{B_0}{6p}}\cosh 6px^{-\frac{B_0}{6p}}(7\sinh 5px + \sinh 7px)$$

and the other eigenfunctions are obtained using the recursive formula

$$\Psi_n(x,B_k) = (-\frac{d}{dx} + W(x,B_k))\Psi_{n-1}(x,B_{k+1}), \tag{45}$$
for $k = 0,\ldots,n-1$, $n = 1,2,\ldots$

For illustration, we will consider the nonnegative potential $V = V_- - (V_-)_{\min}$, with zero-minimum. So,

$$V(x) = V_-(x,B_0) + \frac{1}{6}(35B_0 + 6p)p \tag{46}$$

The corresponding energies are denoted by $E_n$.
$$E_n = E_n^{(-)} - (V_-)_{\min}$$
$$= \left(\frac{5B_0}{6} + p\right)^2 - \left(\frac{5B_0}{2} + p - 7np\right)^2 + \frac{1}{6}(35B_0 + 6p)p \tag{47}$$

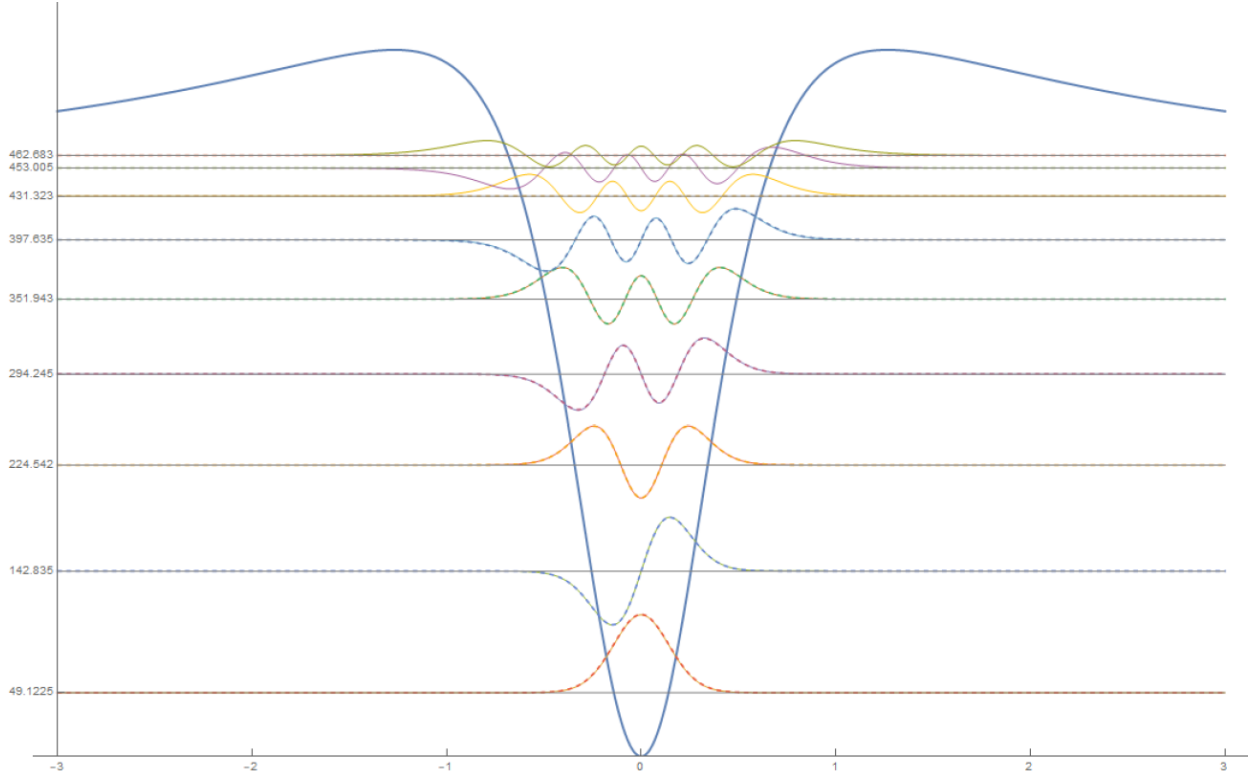

Figure 1. Exact eigenfunctions (solid) and approximated eigenfunctions (dashed) of the potential having superpotential W(x,A,B) =Atanhpx+Btanh6px, where the approximated eigenvalues obtained by NDEigensystems were replaced by the correct and exact eigenvalues for B0=24, p=0.35. We notice that all approximated eigenfunctions match perfectly the exact solutions except for those three states around the four turning points.

The bound state $n$ for the nonnegative potential $V$ need to satisfy the physical constraints on the energies $E_n$ (47): for all $n \geq 1$

$$0 \leq E_{n-1} < E_n \tag{48}$$

We can show that if $0 < p < \frac{B_0}{3}$, then

$$1 \leq n < \frac{5B_0 + 27p}{42p} \tag{49}$$

It is worth mentioning that for small parameter $p$, the potential tends to be biquadratic potential.

Of course, we will need to check that these states are normalizable. By inspecting the eigenfunctions $\Psi_n(x)$, we can see that

$$\Psi_n(x) \propto \cosh px^{-1+\frac{B_0}{6p}} \cosh 6px^{-\frac{B_0}{6p}} P_n(x) \tag{50}$$

where $P_n$ is a real hyperbolic polynomial of the form

$$P_n(x) = \sum_{k=0}^{7n} \alpha_k \cosh(kpx) + \beta_k \sinh(kpx),$$

where the coefficients $\alpha_k, \beta_k$ are constants with at least $\alpha_n \neq 0$ or $\beta_n \neq 0$. The boundary conditions at infinity requires that the limit should be equal to zero. But

$$\lim_{x \to \pm\infty} \Psi_n(x) \propto \lim_{x \to \pm\infty} e^{(-p - \frac{5B_0}{6} + 7np)x}. \tag{51}$$

So, in order for this limit to converge to zero, we impose the condition

$(-p - \frac{5B_0}{6} + 7np) < 0$, i.e.
$$1 \leq n \leq n_{max} \tag{52}$$
where
$$n_{max} = \begin{cases} \left\lfloor \left|\frac{5B_0 + 6p}{42p}\right| \right\rfloor & if \quad \frac{5B_0 + 6p}{42p} \text{ is not an } integer \\ \left\lfloor \left|\frac{5B_0 + 6p}{42p}\right| \right\rfloor - 1 & if \quad \frac{5B_0 + 6p}{42p} \text{ is an } integer \end{cases}$$

and $\lfloor . \rfloor$ represents the floor function, which gives the integer part of the argument. Here, the positive integer $n_{max}$ is the maximum number of bound states allowed by the two physical constraints. So, maybe not all finite bound states allowed in (49) are normalizable, since the difference between the two upper bounds is exactly $\frac{1}{2}$. We might find up to one state to be removed for non normalizability.

In figure 1, we display the plot of the nonnegative potential $V$ with its discrete energies and eigenfunctions for $B_0 = 24$, $p = 0.4$, and in this case, according to (52) the maximum number of bound states is $n_{max} = 7$.

Such potential could be useful to model molecular structure since it has a limited number of excited states.

These potentials with humps could be of great interest to modeling nuclear potential where the strong nuclear force and the electromagnetic force are competing. The potential takes into account forces that are both attractive and repulsive.

Among the solvable potentials mentioned above, the well-known Scarf II potential is the only potential with one hump, see figure 2. We have used the same technique of SUSY shape invariance method in study of the Scarf II with the superpotential being $W^S(x, A, B) = A\tanh px + B\text{sech} px$, and potential partner $V^S$ defined by
$$V^S(x) = A^2 + (sechpx)^2(B^2 - A(A+p) + B(2A+p)\sinh px). \tag{53}$$

Its energy formula is given by $E_n^S = A^2 - (A - np)^2$. The maximum number possible bound states for this model, since its energy formula is a quadratic polynomial in $n$, cannot exceed $\frac{A}{p} + \frac{1}{2}$. For these bounded states, we can further lower this upper limit by taking in consideration the normalizability condition.

When evaluating the eigenfunctions $\phi_n^S$, we can show that there is a pattern similar to our last observation in the previous proposed potential, and that $\phi_n^S$, $n = 0,1,2,...$ satisfy

$$\phi_n^S(x) \propto e^{-\frac{2B\arctan(\tanh\frac{px}{2})}{p}} \cosh px^{-\frac{A}{p}} Q_n(x), \tag{54}$$
where $Q_n$ is a real hyperbolic polynomial in the form
$$Q_n(x) = \sum_{k=0}^{n} \gamma_k \cosh kpx + \eta_k \sinh kpx,$$
where, the coefficients $\gamma_k, \eta_k$, are constants with at least $\eta_n \neq 0$, or $\gamma_n \neq 0$. So,
$$\lim_{x \to \pm\infty} \phi_n(x) \propto \lim_{x \to \pm\infty} e^{-\frac{2B\arctan(\tanh\frac{px}{2})}{p} + (-A+np)x}.$$
Therefore, for the eigenfunctions to satisfy the normalization condition, and since the function arctan is bounded, we must have $1 \leq n < \frac{A}{p}$. To the best of our knowledge, these observations and properties of the Scarf II potential are not available in the literature.

In figure 2, we set the parameters as $A = 7$, $B = 4$, and $p = 1$, and plotted the shifted Scarf II potential with its zero-minimum, eigenvalues and corresponding eigenfunctions. In this case, the maximum allowed bound states $n_{\max}$ is 6. We have included the plot of the numerical solutions using NDEigensystem (NDE), alongside the exact solutions using the recursive formula (45).

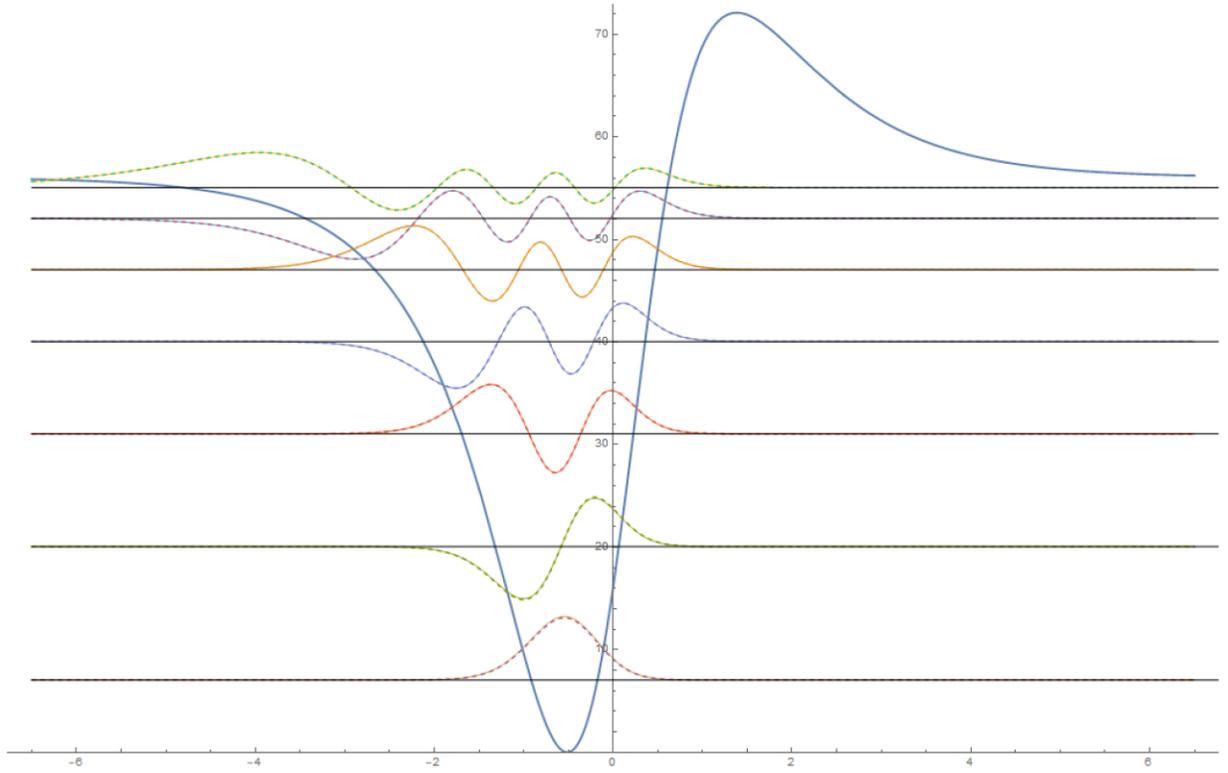

Figure 2. Plot of Scarf II potential $V[x]=A^2+\text{Sech}[p\ x]^2\ (B^2-A\ (A+p)+B\ (2\ A+p)\ \text{Sinh}[p\ x])$, with energy $E_n=A^2-(A-n\ p)^2$. We have shifted the potential. The NDEigensystem numerical (dashed) eigenvalyes and eigenfunctions are matching our exact formulas.

## 5 Comparing Numerical Methods With Exact Solutions

All the results above are in closed form, and we compute the eigenfunctions recursively as we move the ladder starting from the ground state to the $n_{\max}^{th}$ state.

We will compare in this section the exact values of the energies given explicitly in closed form by the formula (17) to the numerical values computed using four approximation methods of these eigenvalues These methods are: the Shooting method, also known as "Wag-the-Dog" method [28], SWKB method, WKB method [15], and with NDEigensystem in Mathematica.

We notice that in this model, the exact eigenvalues of the system are all below the horizontal asymptote of the potential, $v = \lim_{x \to \pm\infty} V_-(x, B_0) = \frac{1}{36}(25B_0^2 + 270B_0 p + 72p^2)$, in this example $v = 463.245$, so, no eigenvalue can reach or exceed this threshold. However, the shooting method and NDEigensystem methods allow states on or above this line and having four turning points.

In table 1, we give the numerical approximations of the eigenvalues and comparing them to the exact energy (47) of the potential $V$ (46). The true relative error $err = \left|\frac{exact-approx}{exact}\right|$ is given as a percentage for each numerical method approximation, relative to the exact value. As it can be seen, except for the ground state and first excited state, all other states below the limit line, have a relative error ranging between 1.093% and 5.09%.

| n | Exact | NDE | err(%) | WKB | err(%) | SWKB | err(%) | Shooting | err(%) |
|---|---|---|---|---|---|---|---|---|---|
| 0 | 49.1225 | 49.1225 | 0 | 50.205 | 2.203 | 49.1265 | 0.008 | 49.1225 | 0 |
| 1 | 142.835 | 142.835 | 0 | 143.826 | 0.693 | 142.839 | 0.002 | 142.84 | 0.003 |
| 2 | 224.542 | 227.456 | 1.297 | 228.356 | 1.698 | 227.464 | 1.301 | 227.456 | 1.297 |
| 3 | 294.245 | 302.920 | 2.948 | 303.730 | 3.223 | 302.936 | 2.953 | 302.92 | 2.948 |
| 4 | 351.943 | 369.135 | 4.884 | 369.859 | 5.090 | 369.163 | 5,090 | 339.62 | 3.501 |
| 5 | 397.635 | - | - | - | - | - | - | - | - |
| 6 | 431.323 | 425.962 | 1.242 | 426.605 | 1.093 | 426.008 | 1.232 | 425.962 | 1.242 |
| 7 | 453.005 | - | - | - | - | - | - | - | - |
| 8 | 462.683 | 463.347 * | 0.143 | - | - | - | - | 463.246 ** | 0.121 |

*Table 1: Potential V energies: exact vs numerical approximation. \* above the asymptote, \*\* on the asymptote.*

In table 1, we have discovered that the numerical methods skipped some of the bound states. In figure 1, we have plotted the eigenfunctions obtained using NDEigensystem (dashed line), with the exact eigenfunction (solid line), and having the same number of nodes, at the same bound state level. We replaced those missing eigenvalues with the exact ones given by (47).

The reason behind this failure to detect some states near the asymptotic line is that; it is well known for example, that near a turning point, WKB and SWKB approximation methods are no more valid. We can observe it in this model, where around the intersection of the $5^{th}$ bound state and the potential $V$, these numerical methods stopped capturing some eigenvalues close to the turning points. Therefore, they failed to account for bound states, starting at that level.

A small error in the numerical approximation can throw the approximate eigenvalues above the horizontal asymptote as can be seen with the NDEigensystem and the shooting methods, which is not allowed by the maximum number of bound state inequality (52). Therefore, instead of having only two turning points if they stayed below the limit line, they end up with four turning points above the line. Hence, it forces the corresponding eigenfunction's numerical approximation to fail beyond the potential's turning points and the outer walls. Some numerical methods try to connect oscillatory classically allowed regions with exponentially classically forbidden regions by patching them smoothly with Airy functions.

Here, we give the NDEigensystem code used in the numerical approximations.
$\mathcal{L} = -u"[x] + V[x]u[x]$ ;
 {vars,funs} $= NDEigensystem[\{, DirichletCondition[u[x] == 0, True]\}$,
  $\{x, -c, c\}, n, Method -> \{"SpatialDiscretization" -> \{"FiniteElement",$
 $\{"MeshOptions" -> \{"MaxCellMeasure" -> 0.001\}\}\},$
 $"Eigensystem" -> \{"Arnoldi", "MaxIterations" -> 10000\}\}];$

## 6 Conclusion

In this paper we have introduced a new family of exactly solvable Schrödinger equations
$$V_-(x, A, B) = -pA(\text{sech}px)^2 - 6pB(\text{sech}6px)^2 + (A\tanh px + B\tanh 6px)^2$$
using the supersymmetric technique, with finite discrete bound states energy in the form
$$E_n^{(-)} = (A + B)^2 - (A + B - 7np)^2,$$
and the corresponding eigenfunctions given recursively.

The proposed potential is obtained by a superpotential formed by two tangent hyperbolic functions, with one being a sextuple angle,
$$W(x, A, B) = A\tanh px + B\tanh 6px.$$
In our previous results, the superpotentials proposed were having double and quadruple angles.

This potential has prospective applications in different fields of physics and chemistry. The potential has a finite number of bound states that can be controlled by varying its parameters. The two humps of this symmetric potential could also be used to model antagonist forces in quantum systems such as nuclear physics and chemistry. The only exactly solvable potential with humps is the Scarf II potential, though it is a one humped-asymmetric potential.

The numerical methods used in the comparison; Shooting, SWKB, WKB methods, and NDEigensystems, are missing some of the bound states determined precisely by our formula.